\documentclass[english, prl,twocolumn]{revtex4}

\usepackage{graphicx}           	% for inserting eps files
\usepackage{amsmath,amsfonts,amssymb}
\usepackage{hyperref}

\graphicspath{{figs/}}		%  path of the images folder 

\usepackage{epstopdf}
\usepackage{amssymb,amsmath}
\usepackage{color}

\usepackage{bm}
\usepackage{bbm}
\usepackage{mathrsfs}

\newcommand{\ie}{$\it{i.e.},~$}

\newcommand{\SI}{Supplementary Information}

\begin{document}

\title{Universality in boundary domain growth by sudden bridging}

\author{ A. A. Saberi$^{1,2}$}
\author{ S. H. Ebrahimnazhad Rahbari$^3$} 
\author{ H. Dashti-Naserabadi$^4$}
\author{ A. Abbasi$^3$}
\author{ Y. S. Cho$^5$}
\author{ J. Nagler$^6$}

\affiliation{$^1$Department of Physics, University of Tehran, P.O. Box 14395-547,Tehran, Iran}
\affiliation{$^2$ School of Physics and Accelerators, Institute for research in Fundamental Science (IPM) P.O. 19395-5531, Tehran, Iran.}
\affiliation{$^3$Department of Physics, Plasma and Condensed Matter Computational Laboratory,  Azarbaijan Shahid Madani University, Tabriz P.O. 53714-161,Iran}
\affiliation{$^4$Physics and Accelerators Research School, NSRTI P.O. 11365-3486, Tehran, Iran}
\affiliation{$^5$Department of Physics and Astronomy, Seoul National University, Seoul 151-747, Korea}
\affiliation{$^6$Computational Physics, IfB, ETH Zurich, Wolfgang-Pauli-Strasse 27, 8093 Zurich, Switzerland}

\begin{abstract}
  We report on universality in boundary domain growth in cluster
  aggregation in the limit of maximum concentration.  Maximal
  concentration means that the diffusivity of the clusters is
  effectively zero and, instead, clusters merge successively in a
  percolation process, which leads to a sudden growth of the boundary
  domains.  For two-dimensional square lattices of linear dimension $L$,
  independent of the models studied here, we find that the maximum of the boundary   interface width, the susceptibility $\chi$, exhibits the scaling
  $\chi \sim L^{\gamma}$ with the universal exponent $\gamma = 1$. The rapid growth of the boundary domain at the percolation threshold,  which is guaranteed to occur for almost {\em any} cluster   percolation process,
  underlies the universal scaling of $\chi$. %universality.
\end{abstract}

\maketitle

%%BEGIN MAIN TEXT
%%%%%%%%%%%%%%%
\textit{\textbf{Introduction.}} Universality is an important concept in 
statistical 
physics which
implies that the critical exponents characterizing the critical
transition do not depend on the microscopic details of the
model~\cite{stanley_1999}. %, kadanoff_2002}. 
%kandanoff not defined by habib!
Percolation on lattices describes the sudden emergence of a spanning cluster together with its fluctuations. %  together with macroscopic connectedness.
In site percolation %, for example,
in euclidean lattices  the order parameter, usually defined as the fraction of occupied sites in the spanning cluster, is studied as a function of the control parameter $p$ (the fraction of occupied sites of the entire lattice).
The percolation %the
universality class is characterized by a given set of critical
exponents that determine the scale invariant behavior immediately
before, precisely at and just after the phase transition from
microscopic to global connectedness~\cite{stauffer, sahimi, saberi2015, nagler2015}. 
%{\color{blue}
However, the scaling and hyperscaling relations leave only two
independent exponents, e.g., $\beta$ and $\nu$ characterizing the
critical behavior of the order parameter and the correlation length
around the critical threshold $p_c$,
respectively, which fully determine the percolation universality
class. The universality, on the other hand, can be encoded by the rich
fractal structure of the percolation clusters at criticality. A
fractal percolation cluster of fractal dimension $D_f=d-\beta/\nu$,
where $d$ is the dimension of the system,
 is composed
of several other fractal substructures including its perimeter (hull),
external perimeter, backbone, and red sites (bonds), etc. For
instance, it is shown~\cite{Coniglio1982}  that
the fractal dimension $d_f^r$ of the red bonds (a red bond is one that
upon cutting leads to a splitting of the cluster) is given by
$d_f^r=1/\nu$ valid in all dimensions $d$ below the critical dimension $d_c$
at which the mean field exponents hold.
Therefore, the universality
can alternatively be given by the fractal geometry of the model in
terms of $D_f$ and $d_f^r$.

In contrast to cluster aggregation processes at low cluster
concentration, boundary domain growth in the limit of maximum
concentration is poorly understood~\cite{ziff_1983, herrmann_1987}. 

Whereas at low concentration a cluster performs a random motion until
it collides with another cluster or the boundary~\cite{meakin_1984,
  meakin_1984_2, family_1985, meakin_1985, cho_2011}, at high concentration the
diffusivity is negligible and the process is well described by
percolation.  Here, we analyze a variety of percolation processes and
ask how a given rule determines the growth of boundary domains.

The simplest of such processes is site percolation which can be
considered a particular model for cluster-size dependent aggregation
at maximal concentration.  To demonstrate the universality of our
framework we study a wide range of models and find that {\em all}
models exhibit the same scaling of the susceptibility
\begin{equation}
  \chi \sim L^{\gamma}
  \label{eq:chi}
\end{equation}
with the exponent, $\gamma \approx 1$. This universality is
remarkable because other observables such as the fractal cluster
dimension and the fractal surface dimension at the percolation
threshold remain model specific.

%%%%%%%%%%%%%% Figure %%%%%%%%%%%%%%%%%%%%%
\begin{figure}[ht]
  \centering
    \includegraphics[scale=0.8]{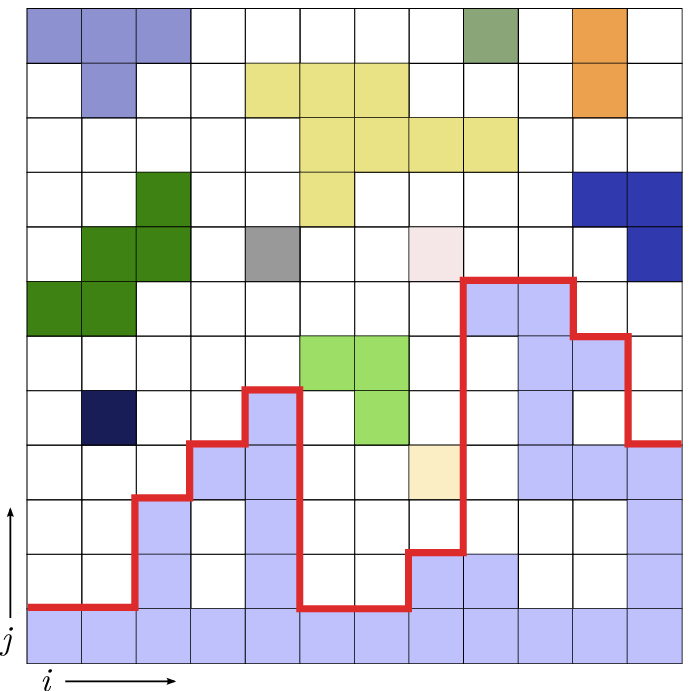}
  \caption{(Color online) {\bf Schematic of the boundary domain.}  The
    (bottom) boundary domain consists of a single cluster (light blue)
    that evolves by merging with other neighboring clusters from the
    initial set of the $L$ bottom sites ($i=0$ to $i=L-1$; $j=0$). The
    red line shows the height profile of the bottom boundary.  Other
    clusters are color coded. White cells are isolated single-site
    clusters (equivalent to unoccupied sites in site percolation).
    \label{fig:boundary}
    }
\end{figure} 
%%%%%%%%%%%%%%%%%%%%%%%%%%%%%%%%%%%%%%%%%%%%%

% MODEL

\textit{\textbf{Results.}} We perform extensive Monte-Carlo simulations of 
cluster 
percolation on
the square lattice with linear dimension $L$.  

Initially all lattice sites represent single clusters of unit size.
Only neighboring clusters 
can merge each time step according to a given rule.
%We repeat this protocol  until a single cluster spans all $N =
%L \times L$ clusters of the lattice.
%, which initially have been placed
%at each lattice site.  
Specifically, choose at each  
time step 
 a cluster and merge the cluster according to a given rule with
one of its neighboring clusters, accessible in its von Neumann neighborhood.
Repeat this over and over again
until a single cluster of size $N$ spans the entire lattice.

During the aggregation process, the system undergoes a phase
transition from a subcritical phase of microscopic $o(N)$-size
components to a supercritical phase with (at least) a macroscopic
component of size $O(N)$. Here we analyze the growth of the lower
boundary domain.  In the beginning, the domain is a single cluster of
size $L$ which merges during the percolation process with other
clusters at its interface, as sketched in Fig.\ \ref{fig:boundary}.

We study models of different universality classes: ({i}) standard
continuous site percolation, and models of discontinuous cluster
percolation ({ii-v}).  (Dis)continuity refers to the behavior of the
order parameter at the critical percolation threshold.  Specifically,
for models ({ii-v}), at each step a cluster is selected uniformly at
random, %meaning 
independent of its size, and ({ii}) merged with its
smallest neighbor cluster, referred to as min-rule, ({iii}) merged
with its largest neighbor cluster (max-rule), or (iv) merged with a
randomly selected neighbor cluster (rnd-rule).  To further demonstrate
the universality of our findings, we also study the recently
introduced ({v}) ''fractional percolation'' rules where the merging of
clusters with substantially different sizes is systematically
suppressed and %. Furthermore, 
components are preferentially merged whose
size ratio is close to a fixed target ratio, $f$.
%, and \hb{ as a
As a result, the order parameter %will 
displays discontinuous %scale-free
  jumps reminiscent of the crackling noise\ \cite{nagler_2013_crack}.  Note
that all these models cover very different aggregation processes in
the limit of maximal density.
%R2.2
Other 'explosive' percolation models, which were proven to be continuous, though exhibiting a substantial gap in the order parameter 
for large finite systems, do not show universality, meaning each microscopic connection rule defines its own universality class.  
Rules (ii)-(v) are truly discontinuous percolation models, 
and thus cannot be related (such as via a set of critical exponents) to standard universality classes of (continuous) percolation.
The main reason why we choose those 'exotic' models is to have a broad spectrum of very different percolation processes.

In the models a neighboring cluster refers to von-Neumann neighborhood
of boundary sites, and cylindrical (half periodic boundary) conditions
are applied.  In order to account for size-dependent delay for
processes (ii)-(v), after each merger, time is advanced by $\delta t =
\min(s_i,s_j)^{1/2}$ where $s_i$ and $s_j$ are the respective relative
sizes of the merging clusters (other choices do not affect any of the
conclusions) \cite{nagler_2013_crack}.

\begin{figure*}[ht]
  \centering
    \includegraphics[scale=0.60]{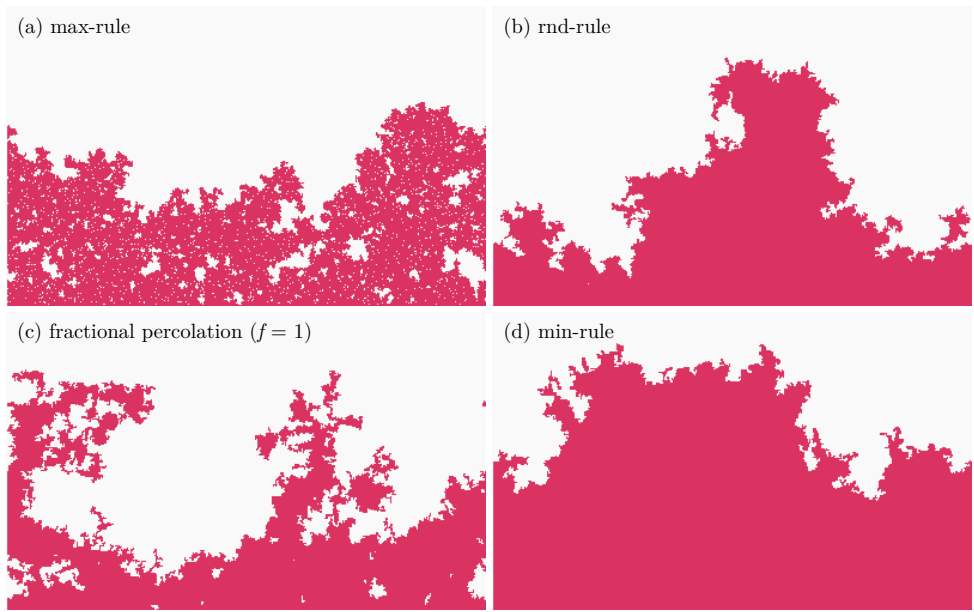}
   \caption{ {\bf Subcritical boundary domains.}  Snapshots of the
    growing boundary domain for different models %at 
    %an intermediate
    exactly one step $\delta t$ before percolation.
    %stage before percolation.  
    (a) max-rule %(a) %and (b) 
    %rnd-rule 
    produces a very porous and loose boundary domain. 
    (b) rnd-rule %(b) 
  %d) Min-rule
    generates a dense and space-filling cluster (fractal dimension
    $D_f=2$). 
    %shows a compact boundary domain.
     (c) fractional percolation
    ($f=1$)  exhibits an almost compact boundary domain.
    (d) min-rule shows a compact boundary domain without voids inside its bulk.
      All models on square lattice of size $400\times 400$; shown is the lower domain of
    size $400 \times 250$.
    \label{fig:clusters}}
\end{figure*}

Typical snapshots of the growing boundary domain for the (a) max-rule,
(b) rnd-rule, (c) fractional, and (d) min-rule demonstrate that the
roughness and the porosity of the boundary are strongly dependent on
the growth mechanism (Fig.\ \ref{fig:clusters}).

%%%%%%%%%%%%%% Figure %%%%%%%%%%%%%%%%%%%%%
%%%%%%%%%%%%%%%%%%%%%%%%%%%%%%%%%%%%%%%%%%%%%

%%%%%%%%%%%%%% Figure %%%%%%%%%%%%%%%%%%%%%
\begin{figure}[ht]
  \centering
    \includegraphics[scale=0.3]{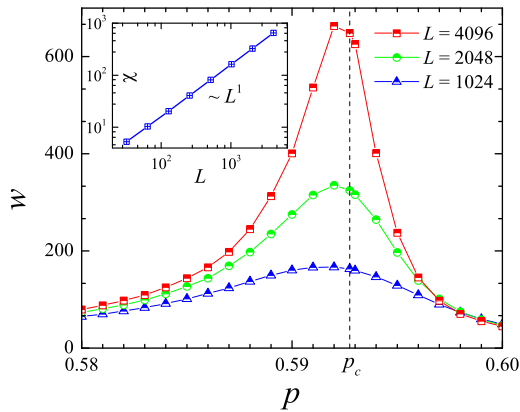}
  \caption{(Color online) {\bf Roughness of the boundary domain in
      site percolation.}  The rms fluctuations of height, $w$, as a
    function of occupation probability $p$ 
    (equivalent to time using
    kinetic formulation, i.e.\ $t=p$ for ordinary percolation).  
    }
    %Square lattice $512\times512$, $10^5$ realizations.
    Inset: Scaling of the susceptibility with size. Square lattices of size $L$, $10^5$ realizations.
    %Inset: Scaling of the susceptibility.
     Error bars are smaller than symbol size.
    \label{fig:w_peak}
\end{figure} 
%%%%%%%%%%%%%%%%%%%%%%%%%%%%%%%%%%%%%%%%%%%%%

For min-rule (ii) and rnd-rule (iv) the critical boundary domain cluster is a compact
surface fractal, meaning the fractal dimension of the boundary
cluster, $D_f = d=2$ (in the thermodynamic limit), where $d$ is the lattice dimension.  This
implies a genuine discontinuity of the percolation phase transition
\cite{herrmann_2010, herrmann_2011}.  For other models, the fractal
dimension of the boundary domain is close to $D_f=91/48\approx 1.89$
characteristic of the standard percolation universality class in two
dimensions (see \SI).  The universality of Eq.\ (\ref{eq:chi})
holds regardless whether or not the critical percolation cluster is
fractal or compact (or whether or not the cluster surface is smooth or
fractal).  To demonstrate this, we study the roughness of the surface
of the growing boundary characterized by the rms (root mean square)
fluctuation of heights, $w$,
\begin{equation}\label{eq:w}
w = \left< \sqrt{\sum_i (h_i - \bar{h})^2 / L} \right>,
\end{equation}
where $h_i$ are the height of the boundary domain at boundary
positions $i=0,\ldots L-1$, and mean $\bar{h}=\sum_i h_i/L$, see
Fig.\ \ref{fig:boundary}.

\begin{figure}
  \centering
    \includegraphics[scale=0.3]{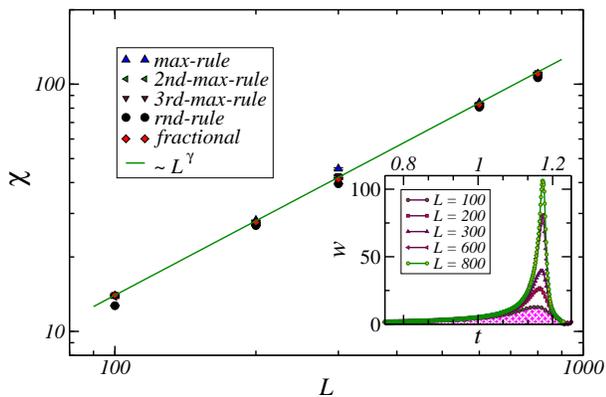}
  \caption{(Color online) \label{fig:rms} {\bf Universality of the
      susceptibility scaling.}  Inset: interface width, $w$
    (Eq.\ (\ref{eq:w})), for rnd-rule as a function of time for
    different lattice size $L$.  Main panel: Maximum of the interface
    width, the susceptibility, Eq.\;(\ref{eq:chi}), at the percolation
    point, as a function the lattice size $L$ for max-rule
    ($\bigtriangleup$), 2nd-max-rule ($\triangleleft$), 3rd-max-rule
    ($\triangledown$), rnd-rule ($\circ$), and fractional ($\diamond$,
    $f=1.0$) .  The solid line shows the best fit, $\chi \sim L
    ^{\gamma}$, where $\gamma \approx 1$ (max-rule: $\gamma=0.95\pm
    0.005$,
    2nd-max-rule: $\gamma=0.98\pm 0.005$, 
    3rd-max-rule: $\gamma=1.00\pm 0.005$, 
    rnd-rule: $\gamma=1.01\pm 0.005$,
    fractional: $\gamma=0.99\pm 0.005$).  
    $800$ realizations for each data point. 
    }
\end{figure}

The rms fluctuation of heights, $w$, exhibits a peak at the
percolation point, $t=t_c$ ($t=T/N$ is the scaled time, 
$T$ denoting the MC steps), as shown for model (i) in
Fig.\ \ref{fig:w_peak}. Most remarkably, the susceptibility, $\chi$,
defined as the maximum of $w$,
\begin{equation}
\chi =  \max_{t}\left[w\left(t\right)\right]
\end{equation}
increases with lattice side length, $\chi \sim L ^{\gamma}$,
independent of the models used here, with the exponent, $\gamma \approx 1$
(Fig.~\ref{fig:rms}).

%R2.4.b:
%Since, as expected,  
In diffusion limited aggregation, for example, microscopic particles diffuse until they touch other particles or the boundary
if in geometrical confinement. Such processes are thus characterized  by a continuous growth of the boundary domain,
where usually $\gamma=1/2$.
In contrast, through rules (i-v) a successive aggregation of clusters that cannot move 
and are initially nearest neighbors is studied. In case (i) this aggregation is known as ordinary site percolation.
Since ordinary site percolation exhibits a rapid but continuous emergence of a unique giant cluster exactly at $p_c$,
and all other clusters are of size $O(\log N)$ before and after $p_c$,
the naive expectation would be that boundary growth may also be continuous (thus characterizing by some $\gamma<1$, if not $\gamma=1/2$).
In addition,  the 
  fractal geometry of the giant percolation cluster and its boundary
 do depend  on the model \cite{sahimi,  herrmann_1987, herrmann_2011} (see Supplementary Information).
So, a universal $\gamma=1$  
%this 
 is a rather surprising finding.
In the following, we explain the
universality by the necessary occurrence of a sudden bridging.

\begin{figure}
  \centering
  \includegraphics[scale=0.3]{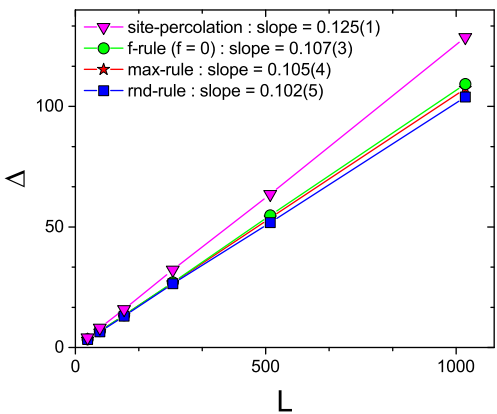}
  \caption{(Color online) {\bf The maximal gap in ${\bf w}$.}  Size
    $\Delta$ of the largest gap in $w$ for a collection of continuous
    and discontinuous cluster percolation models.  Specifically, for
    rnd-rule ($\circ$), 2nd-max-rule ($\square$), 3rd-max-rule
    ($\diamond$), fractional ($\bigtriangleup$, $f=0.5$), all yielding
    discontinuous percolation, and max-max-rule (select at random a
    cluster and merge the two largest clusters that are neighbors of
    each other among the selected cluster and all its neighbors),
    yielding continuous percolation, $\Delta$ as a function of lattice
    size $L$ is shown. $800$ realizations for each data point.  Error
    bars are smaller than symbol size.
    \label{fig:gap}
  }
\end{figure}

Consider the largest single step jump in $w$, 
\begin{equation}
\Delta:=\max_i \left[ (w(t_{i+1})-w(t_i) \right],
\end{equation} 
which occurs at the percolation point, $t_c$, as shown in Fig.~\ref{fig:gap}. 

{\color{black} Because the spanning cluster is macroscopic, \ie of
  size $O(N)$, the linear dimension of the percolation cluster is of
  size $O(L)$, in {\em any} linear dimension. 
  Thus, at percolation an $O(L)$
  number of boundary sites $h_i$ jump from $o(L)$ to $O(L)$.
\paragraph{\bf Case 1:} 
$\alpha L$ number of sites $h_i$ increase to $O(L)$, and $(1-\alpha)
L$ sites stay of size $o(L)$, with some $0<\alpha<1$. Then the mean
difference $\langle h_i-\bar{h} \rangle$ is of size $O(L)$, resulting
from the $(1-\alpha) L$ fraction of sites that have an $O(L)$-sized
difference to $\bar{h}$.  Thus $w^2=o(L)^2 \rightarrow O(L)^2$.
\paragraph{\bf Case 2:} 
Assume $\alpha=1$, meaning all sites jump from $o(L)$ to $O(L)$.
Unless the spanning cluster exhibits only $o(L)$ fluctuations parallel
to the boundary domain, the mean difference $\langle h_i-\bar{h}
\rangle$ is of size $O(L)$. Recall that %, because the linear
the linear dimension of the percolation cluster is of size $O(L)$ in {\em any}
direction, in particular parallel to the height profile $h_i$. Thus
$w^2=o(L)^2 \rightarrow O(L)^2$.
 \paragraph{\bf Case 3:} 
All sites jump from $o(L)$ to $O(L)$ {\em and} the height profile of
the spanning cluster exhibits only $o(L)$ fluctuations parallel to the
boundary domain. In this case, $\langle h_i-\bar{h} \rangle=o(L)$,
with a possible $o(L)$ fraction of sites that show variations of size
$O(L)$.  This determines the spanning cluster not only necessarily
compact (characteristic of discontinuous percolation) but rectangular
(possibly with ''micro-cracks``). This very special case does not show
a macroscopic jump in $w$.
 
We conclude that bridging implies $w^2=o(L)^2 \rightarrow O(L)^2$ and
thus $\chi=\max_t \left[w(t)\right] \sim L^\gamma$ with $\gamma=1$,
virtually independent of the model.}

%%%%%%%%%%%%%% Figure %%%%%%%%%%%%%%%%%%%%%
%%%%%%%%%%%%%%%%%%%%%%%%%%%%%%%%%%%%%%%%%%%%%

%%%%%%%%%%%%%% Figure %%%%%%%%%%%%%%%%%%%%%
%%%%%%%%%%%%%%%%%%%%%%%%%%%%%%%%%%%%%%%%%%%%%

%%%%%%%%%%%%%% Figure %%%%%%%%%%%%%%%%%%%%%
\begin{figure}
  \centering \includegraphics[scale=0.6]{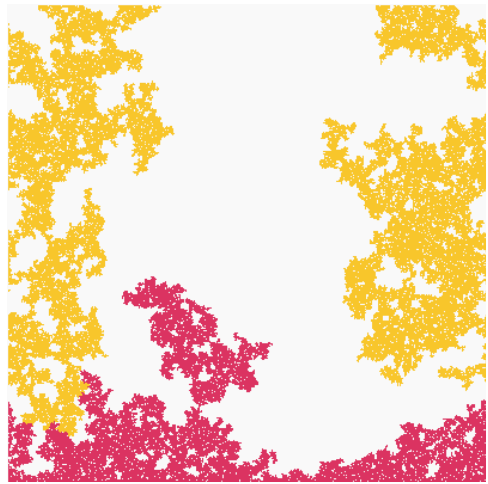}
  \caption{(Color online) 
    \label{fig:illustration}
    {\bf Sudden bridging.}  The fractal boundary domain (bottom, red)
    suddenly gets connected to the spanning cluster (yellow).  This
    sudden event represents case (3) and induces a discontinuity in the domain growth leading
    to $\gamma=1$.}
\end{figure} 
%%%%%%%%%%%%%%%%%%%%%%%%%%%%%%%%%%%%%%%%%%%%%

\textit{\textbf{Discussion.}} In continuous percolation, the emergence of a 
unique 
macroscopic
cluster necessarily coincides with the occurrence of spanning (when
facing sides of the lattice get connected by a path of sites). At the
percolation threshold the giant component is fractal and spanning.
Discontinuous percolation, however, can show a much richer dynamics
than case (3) \cite{herrmann_2010, herrmann_2011, riordan_2011,
  nagler_2012, nagler_2013_phase,  ChenPRE2013a, cho_2013, ChenPRL2014}.  In discontinuous percolation
the emergence of a macroscopic cluster must not necessarily coincide
with the emergence of a spanning cluster, nor must the giant component
be unique at percolation.  Instead, multiple giant (compact)
components can emerge simultaneously \cite{krapiv_2005, souza_2011},
which may merge in multiple discontinuous transitions. Spanning can
occur much later than the first emergence of the macroscopic
component.  Nevertheless, as illustrated in
Fig.\ \ref{fig:illustration}, there necessarily occurs a single event
where one of the $O(N)$-size components connects to the boundary
domain, yielding an $O(L)$-size jump in $w$.  This predicts $\gamma=1$
for both continuous and discontinuous processes, except for very
particular processes.

We call those processes {\em needle growth processes}: With sufficient
preference choose mergers such that the aspect ratio of the cluster
that results from the merging is as large as possible.  This rule (and
other artificially constructed rules) would lead to spanning prior to
the emergence of a macroscopic cluster.  Thus the boundary domain
would increase continuously in the thermodynamic limit.

Notably, processes where the emergence of a macroscopic cluster
proceeds spanning are also possible: With sufficiently large
preference grow the second largest cluster in the system such that its
aspect ratio stays as close as possible to unity. This guarantees the
simultaneous emergence of two macroscopic ($O(N)$-size) compact
clusters reluctant to span the lattice at the percolation threshold
(defined via the first emergence of a macroscopic cluster and not via
spanning).

However, boundary domain growth for those processes would still
exhibit an $O(L)$-size jump in $w$ because spanning is certain at
times during the process.

Continuous domain growth is not only expected for {\em needle
  processes} but known for a broad class of physical relevant
processes.  
Examples include, percolation or aggregation processes
where boundary growth is the dominating process such as in invasion
percolation %.
%, or in cluster aggregation processes at {\em non-maximal}
%concentration where clusters have a sufficient freedom to perform
%random walks \cite{cho_2011}.  
%Examples include 
or KPZ growth models \cite{kardar_1986}.
 % diffusion-limited and cluster-cluster aggregation
 % processes \cite{meakin_1984, meakin_1984_2, family_1985,
 %   meakin_1985, kardar_1986}.
%
More specifically, 
classification of the evolution of (1+1)-dimensional boundary domains in non-equilibrium growth processes has been very well established in the past \cite{stanley_1995}.
% [ A.-L. Barab\'{a}si, H.E. Stanley, Fractal Concepts in Surface Growth, Cambridge University %Press, Cambridge, 1995]. 
% 
 One of the most important universality classes is given by the Kardar-Parisi-Zhang (KPZ) 
 \cite{kardar_1986}
 %[ [*] M. Kardar, G. Parisi, Yi-C. Zhang, Phys. Rev. Lett. 56, 889 (1986)] 
 equation $\partial h(x, t)/\partial t = \nu \nabla^2h + \lambda \left\vert \nabla h \right\vert ^\mu + \eta(x, t)$ with $\mu=2$, which also includes the Edwards-Wilkinson (EW) universality for $\lambda=0$. The boundary fluctuations reach a maximum $\chi$ in the stationary state which scales with the system size as $\chi\sim L^\gamma$. It is shown \cite{amar_1993}
 %[ [**] J.G. Amar and F. Family,
%Phys. Rev. E 47, 1595 (1993)] 
that in the presence of the additive noise $\eta$, the roughness exponent $\gamma$ falls into the ordinary KPZ (EW) class with the exact value  \cite{kardar_1986} 
$\gamma=1/2$ for all $\mu$ ($\lambda=0$). 
However, for the deterministic case of $\eta=0$ and for $\mu<1$, an instability occurs which leads to a fluctuating grooved interface. In this case, the roughness exponent is observed
to coincide with our prediction $\gamma\approx 1$ %in the limit of large system sizes 
\cite{amar_1993}. 

To conclude, boundary domain growth at maximal concentration is
discontinuous and characterized by a universal exponent with respect
to the scaling of the maximum of the boundary interface width.  The
universality for boundary domain growth at maximal concentration in
terms of the model-independent exponent $\gamma=1$ is explained by the
necessary occurrence of sudden bridging, the connection of the
boundary domain to the largest cluster in the system. Our
  study opens a new category of growing interfaces complementary to
  the well-established self-affine surfaces. Loosely speaking, in the
  non-isotropic {\em self-affine} growing interfaces, the exponent
  $\gamma$, 
  which determines the universality class of the growth process,
  is model specific
  % and determines the universality of the
  %growth process, 
  while the fractal properties of the boundary domain
  and its surface (if any) do not have any information about the
  universality class. In our case, the story is rather inverse: for an
  isotropic {\em self-similar} growing interface, the fractal
  structure is model specific characterizing the universality classes
  (if any), while the exponent $\gamma=1$ is super-universal for all
  models. In this picture, the exponent $\gamma$ captures the
  underlying isotropic symmetry in the growth processes.

We found a universal scaling behavior of an important observable across a wide range of percolation models (i.e., for discontinuous and continuous percolation) 
that has not been reported as of yet:
 a universal scaling of the boundary domain growth induced by a phenomenon which we call sudden bridging.
Previous aggregation models (i.e., diffusion limited aggregation) assume
that microscopic particles diffuse until they collide with other particles (or the boundary), which usually leads to $\gamma=1/2$ (and not $\gamma=1$).
In a broader context 
as an empirical application of our finding, it is worth noticing that one of the crucial aims in surface growth science is to devise a dynamical growth model and mechanisms to understand the underlying physics behind the observed height profile in the lab using different tools, e.g., Atomic Force Microscopy (AFM). In AFM sample scans, the tip which moves along a 1d sample, only sees effective columnar valleys regardless of the inherent complex fractal structure of the grown surface a little deep inside. 
In this light,
 our study suggests that different percolation-based growth processes with different characteristic complex inherent structures can lead to the same statistics observed at the effective surface of the samples. To our knowledge, such correspondence has never been reported yet.

\textit{\textbf{Methods.}} We perform large scale Monte-Carlo simulations on a 
$2D$ 
square
lattice of length $L$. Periodic boundary conditions are applied along the horizontal
$x$-direction. We start with $N = L \times L$ single clusters (meaning at $t=0$
each site represents an individual cluster).

At each time step, we merge two neighboring clusters according to a fixed 
rule. We choose von Neumann neighborhood (i.e.\ given by either $x\pm 1$, or $y\pm 1$;
sites or clusters with a double displacement in $x$ and in $y$ direction
are no neighbors).
%to define the vicinity of the neighboring clusters.

%Therefore, 
At each MC step the number of clusters in the system
decreases by 1 and eventually at the end of simulations one cluster
emerges which then covers the entire lattice.

We use two different markers %colors on the grid 
to identify bulk and domain clusters. 
At the beginning of the simulations, all the sites (and %, or
%equivalently 
clusters) of the first row of the grid (at $y = 0$) are
marked black while the rest are white. Hence, 
%The former represents the
the boundary domain at $t = 0$ constitutes of $L$ clusters at $y=0$
whereas the bulk constitutes $L\times (L-1)$ clusters in the domain $y\ge 1$.
%, and the latter serves as clusters at the
%bulk. 
Whenever a %white (
bulk %) 
cluster (marked in white) is merged with a %black (
domain %)
cluster (marked in black) it will join the domain. %s color changes to black. 
%Therefore, black is the dominant
%color which introduces the bias in the system. 

As time advances, the
interface at the bottom will experience an upward directed but stochastic growth. The percolation time, $t_c$, is defined through the MC step at which the boundary domain  touches the ceiling of the system (at $y=L-1$), usually referred to as spanning.

Except for standard site percolation, we study two models types:
%Growth process is based on two types of kernels: 
(i) {\em Focal models}: %and (ii) {\em non-focal models}. 
For focal models we choose randomly a cluster {\em (focal cluster)} and merge
it with one of its von Neumann neighboring {\em clusters}.
Specifically,  max-rule means choose at random a  cluster and merge it with the largest neighboring cluster,
min-rule  means choose at random a cluster and merge it with the smallest neighboring cluster and
for  rnd-rule we choose at random a  cluster and merge it with randomly chosen neighboring cluster.
For the {\em fractional rule} 
choose at random a  cluster and merge it with the von Neumann neighboring cluster (nn) that minimizes $f s_f - s_{nn}$ 
where $s_f$ is the size of the focal cluster,
$s_{\text{nn}}$ the size of the neighboring cluster and
$f$ a constant (a parameter of the model that controls to what extent clusters of a certain size ratio merge preferentially together \cite{nagler_2013_crack}).
Models based on focal kernels thus necessary involve the growth of the randomly chosen focal cluster.

In contrast, we study also (ii) {\em non-focal models} where the focal cluster does not necessarily aggregate 
with some other at a given MC step: 
choose at random a cluster, independently of its size (call this cluster the {\em focal cluster}).
The focal cluster will be surrounded by other clusters (call these clusters {\em neighboring clusters}),
 which share at least a single von Neumann neighboring site (i.e.\ coordinate displacements $x\pm 1$, or $y\pm1$ define the neighboring).
Now consider the set $S:=\{\text{focal cluster}, \text{all neighboring clusters}\}$.
Merge 
%among the collection of
%the focal cluster and all its neighbor clusters 
two clusters of the set $S$ that are neighboring, according to some given fixed rule
 (merging two clusters in $S$ that are not neighbors is forbidden).
Specifically,
max-max rule: choose at random a cluster, call it focal cluster, find  the largest cluster
 in the set {\em focal cluster plus all von Neumann neighbors} and merge it with its largest neighbor cluster.
2nd-max rule: choose at random a cluster, find the second largest cluster
 in the set {\em focal cluster plus all von Neumann neighbors} and merge this cluster with its largest neighbor.
3rd-max rule: choose at random a cluster,  find the third largest cluster in the set {\em focal cluster plus all von Neumann neighbors}
and merge this cluster with its largest neighbor.

\textit{\textbf{Acknowledgments.}}  
A.A.S. and H.D.N. acknowledge partial financial supports by the Iran National Science Foundation (INSF), and University of Tehran. 
J.N. thanks the ETH Risk Center for support and L. IIie-Deustch for invaluable input.

\section*{Author contributions statement}

A.A.S., S.H.E.R. and J.N. have equally contributed in designing the research. 
H.D.N. and S.H.E.R did the simulations. A.A.S. and J.N. conceived and analysed 
the data. A.A.S. wrote the supplementary paper whose simulations were done by 
H.D-N.. J.N., A.A.S and S.H.E.R wrote the manuscript. A.A. and Y.S.C.had early 
contribution in the work.

\section{Supplementary information}

 In this supplementary material, we will present some additional details of 
 simulations and the results reported in the paper. It includes the study of 
 the ordinary (site and bond) percolation problems together with the details of 
 computations for the scaling properties of critical clusters.

 \subsection{Standard Percolation Models}
 
\begin{figure}
 \centering
 \includegraphics[scale=0.8]{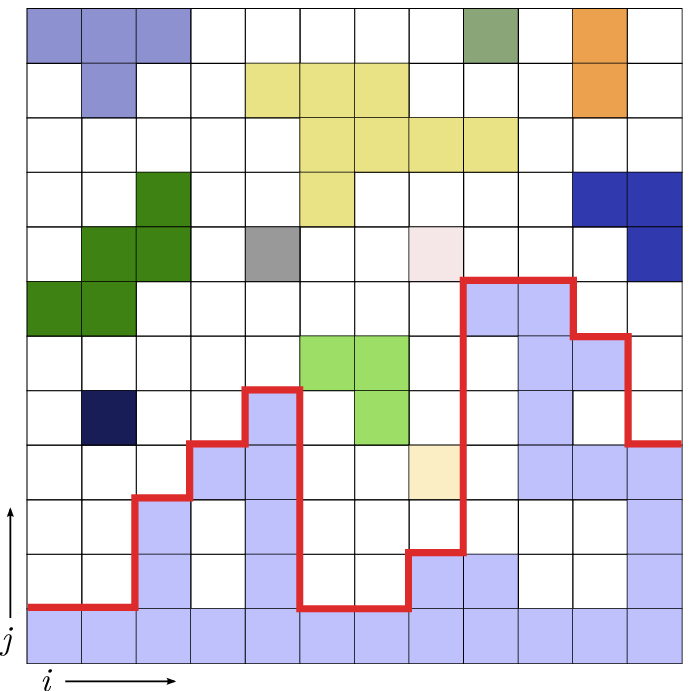}
 \caption{The boundary domain is the cluster of occupied sites attached to the 
 bottom boundary sites which are fixed to be occupied as a boundary condition. 
 The solid line shows the height profile attributed to each boundary site 
 ($i,j=1$).}
 \label{fig:per}
 \end{figure}

\begin{figure}
 \centering
 \includegraphics[scale=0.39]{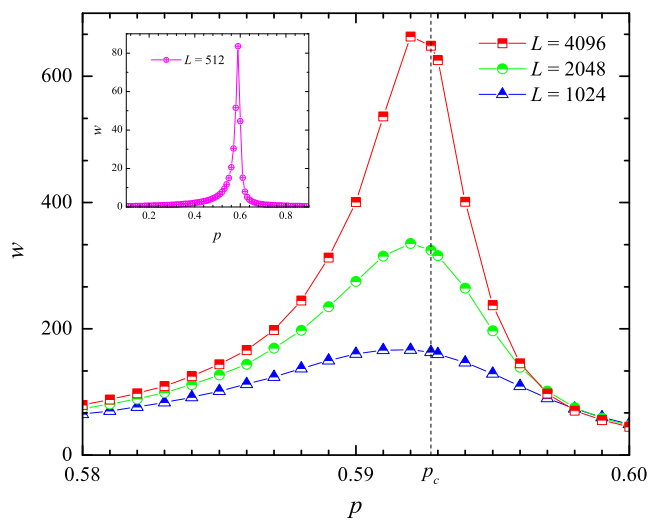}
 \caption{Main: The width $w$ as a function of the occupancy $p$ around $p_c$, 
 for different system sizes $L$. Inset: $w$ in the whole interval $p\in(0,1)$. }
 \label{fig:w_p}
 \end{figure}

\begin{figure}
 \centering
 \includegraphics[scale=0.4]{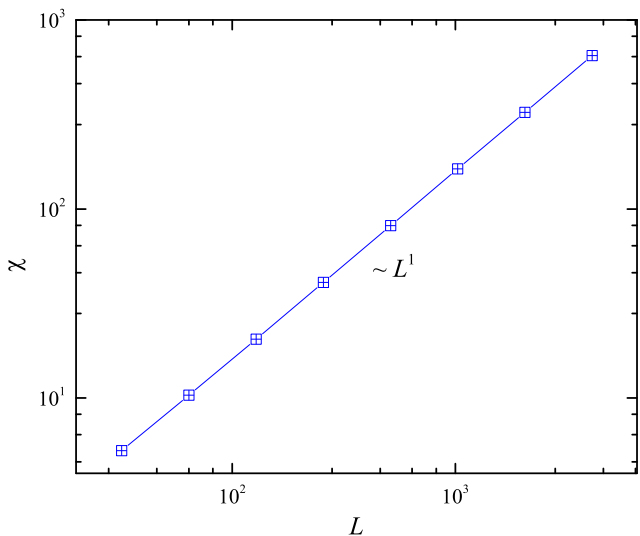}
 \caption{For the ordinary site percolation model, the susceptibility $\chi$ 
 shows a 
 scaling relation with the system size $L$, i.e., $\chi \sim L^\gamma$, with the 
 exponent $\gamma$ very close to $1$.}
 \label{fig:chi_L}
 \end{figure}

 In order to examine our computations for the ordinary percolation with a 
 continuous phase transition, we first 
 consider the site percolation model on a square lattice of different sizes 
 $L=2^k, k =\{5,6,\dots, 12\}$. Each site can be either in an occupied or 
 unoccupied state with probability $p$ or $1-p$, respectively. All 
 nearest-neighbor occupied sites will define a cluster assigned by a specified 
 color. As 
 a boundary condition, we fix all sites at the bottom boundary ($i,j=1$) in an 
 occupied state which will be intact in time and constitutes the boundary domain 
 (the cluster with the same color as the bottom-boundary occupied sites) whose 
 statistical evolution is our main point of interest here. More precisely, to 
 each boundary site ($i,j=1$) we attribute a height function $h_i$ which is the 
 maximum height of the occupied site in the column  $i=1,2,\cdots,L$ 
 belonging to 
 the boundary 
 domain (see Fig. \ref{fig:per}). By running the occupancy $p$ from $0$ to $1$, 
 some occupied sites will randomly join to the boundary domain and thus the 
 height profile 
 $\{h_i\}$ will evolve as a function of $p$.     
 The first quantity of interest is the height fluctuations measured by the root 
 mean square (rms) $w$ of the height profiles 
 
 \begin{equation}
 \langle w \rangle_E = \left\langle \sqrt{\sum_{i}(h_i-\bar{h})^2 / L} \right\rangle_E,
 \end{equation}
 where $\bar{h}$ is the mean height and $\langle \cdots \rangle_E$ denotes for  
 ensemble 
 averaging. For a given occupancy $p$ and system size $L$, the averages are 
 taken over more than 
 $5000$ independent samples. We find 
 that the width $w$ exhibits a peak whose position converges to the site 
 percolation threshold 
 $p_c=0.5927\dots$ for large system sizes (see Fig.\ \ref{fig:w_p}). At $p=p_c$, 
 the boundary domain spans the lattice along the vertical direction. We also 
 find 
 that the value $\chi = w(p_c)$, which is called susceptibility, 
 exhibits a scaling relation with the system size as $\chi \sim L^\gamma$. To 
 estimate the exponent $\gamma$, the value of $\chi$ is averaged over $5\times 
 10^4$ samples for each $L$, and we find that $\gamma = 0.997(3)$ 
 (Fig.\ \ref{fig:chi_L}), very close to 
 $1$ in accord to the corresponding exponent for other percolation models even 
 with discontinuous phase transition. We find the same exponent $\gamma\sim 1$ 
 for the bond percolation model as well.
 
 
 \subsection{Cluster Statistics}
 
\begin{figure}
 \centering
 \includegraphics[scale=0.37]{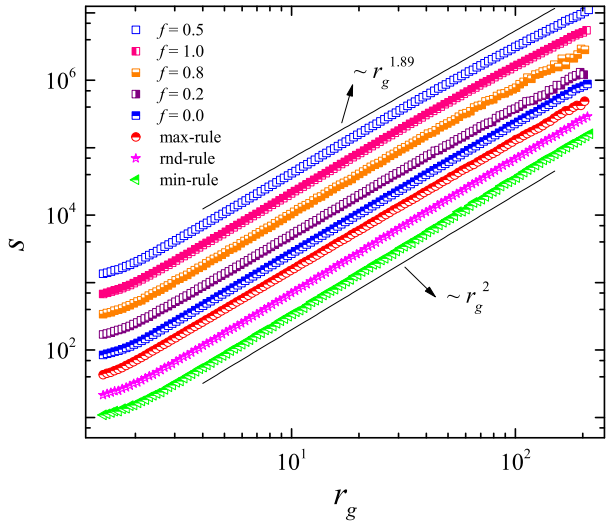}
 \caption{The average size $s$ of critical clusters versus their average radius of 
 gyration $r_g$ for different rules obtained by averaging over $10^4$ 
 independent samples of size $L=1024$.}
 \label{fig:s_rg}
 \end{figure}

\begin{figure}
 \centering
 \includegraphics[scale=0.34]{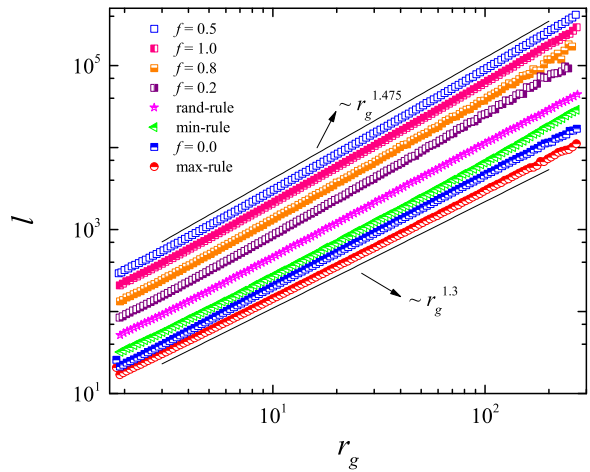}
 \caption{The average length $l$ of the cluster boundaries versus their average 
 radius of gyration $r_g$ for different rules obtained by averaging over $10^4$ 
 independent samples of size $L=1024$.}
 \label{fig:l_rg}
 \end{figure}

 In this section we present the results of our computations for the critical
 clusters 
 of different rule models including the min-rule, max-rule, rnd-rule and the 
 class of 
 fractional percolation rules i.e., $f$-rule for $0\leq f\leq 1$.
 The fractal dimensions of the critical clusters and their boundaries (or loops) 
 are measured by examining the scaling relation between the average size $s$ of 
 the clusters, and the average length $l$ of their boundaries with their average 
 radius of gyration $r_g$, respectively (i.e., $s\sim r_g^{d_c}$ and $l\sim 
 r_g^{d_l}$ where $d_c$ and $d_l$ denote for the fractal dimension of a critical
 cluster or its boundary, respectively--Figs. \ref{fig:s_rg} and 
 \ref{fig:l_rg}). 
 
  
\begin{figure}
 \centering
 \includegraphics[scale=0.36]{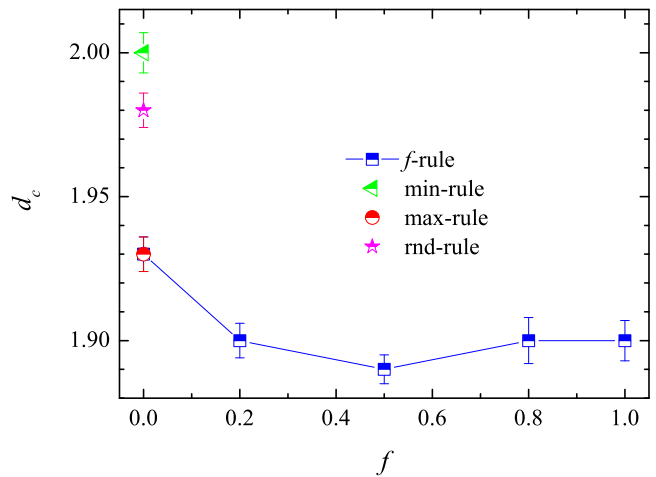}
 \caption{The fractal dimension $d_c$ of a critical cluster for different rules.}
 \label{fig:dc}
 \end{figure}

\begin{figure}
 \centering
 \includegraphics[scale=0.36]{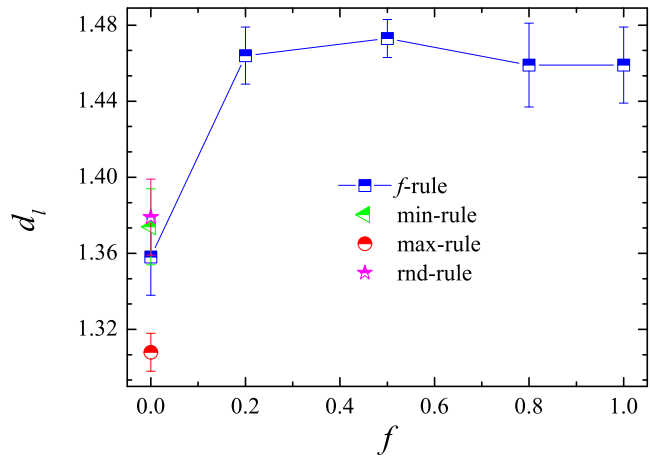}
 \caption{The fractal dimension $d_l$ of a critical
  cluster boundary for 
 different rules.}
 \label{fig:dl}
 \end{figure}

 Figures \ref{fig:dc} and \ref{fig:dl} summarize the values of computed fractal 
 dimensions for the critical clusters and their boundaries, respectively. Among 
 them, the 
 min-rule gives rise to compact clusters of dimension $2$ with fractal 
 boundaries while for the other rules the critical clusters seem to have a porous 
 structure. The other characteristic feature is that for different $f$-rules 
 with $f>0$, both fractal dimensions $d_c$ and $d_l$ seem to be $f$-independent 
 within the error bars.

 %We have also examined the scaling relation $n \sim s^{-\tau}$, where
 %$n$ is the overall cluster size distribution at criticality for
 %different rules. We find a conclusive scaling behavior only for the
 %max-rule and $f$-rule with $f=0$. The results are summarized in Table
 %\ref{tab:tau_s}.
 
 We have also examined the scaling relation $n_s \sim s^{-\tau}$, where
 $n_s$ is the void size distribution in the spanning cluster at
 criticality for different rules. We find a conclusive scaling behavior
 only for the max-rule and $f$-rule with $f = 0$. For the other models,
 the spanning clusters are so compact either without or with a little
 average number of voids inside which elude a power-law behavior. The
 results are summarized in Table\ \ref{tab:tau_s}.

 \begin{table}
 \centering
 \caption{The exponent $\tau$ for different rule models.}
 \begin{tabular}{|c||c|c|}
 
 \hline rule & max & $f=0$ \\
 \hline  $\tau$& $1.79(2)$ & $1.75(2)$ \\ 
 \hline 
 \end{tabular} 
 \label{tab:tau_s}
 \end{table}

\end{document}